\documentclass{article}
\usepackage{iclr2019_conference,times}

\usepackage{amsmath,amsfonts,bm}

\def\eqref#1{equation~\ref{#1}}

\def\1{\bm{1}}

\DeclareMathAlphabet{\mathsfit}{\encodingdefault}{\sfdefault}{m}{sl}
\SetMathAlphabet{\mathsfit}{bold}{\encodingdefault}{\sfdefault}{bx}{n}

\newcommand{\E}{\mathbb{E}}

\newcommand{\Var}{\mathrm{Var}}

\usepackage[utf8]{inputenc} 
\usepackage[T1]{fontenc}    
\usepackage[hidelinks]{hyperref}

\usepackage{url}            
\usepackage{booktabs}       
\usepackage{amsfonts}       
\usepackage{amssymb}
\usepackage{nicefrac}       
\usepackage{microtype}      
\usepackage{bbm}

\usepackage{color}
\usepackage{amsmath}
\usepackage{amsthm}
\usepackage{graphicx}
\usepackage{subcaption}
\usepackage{tikz}
\usepackage{caption}
\usepackage{multirow}
\usepackage{wrapfig}
\usepackage{algorithmic}
\usepackage{algorithm}
\usepackage{amsmath}
\usepackage{soul}
\usepackage[normalem]{ulem}
\usepackage{bm}
\usepackage[title]{appendix}
\usepackage{hyphenat}
\usepackage{gensymb}
\usepackage{array}
\usepackage{makecell}

\title{A Deep Probabilistic Model for Customer\\Lifetime Value Prediction}

\author{Xiaojing Wang\\
	Google\\
	Mountain View, USA\\
	\texttt{xiaojingw@google.com}\\
	\And
	Tianqi Liu\\
	Google Research\\
	New York, USA\\
	\texttt{tianqiliu@google.com}\\
	\And
	Jingang Miao\\
	Google\\
	New York, USA\\
	\texttt{jmiao@google.com}\\
}

\newcommand{\lognormal}{\operatorname{Lognormal}}

\iclrfinalcopy
\begin{document}

\maketitle

\begin{abstract}

Accurate predictions of customers' future lifetime value (LTV) given their attributes and past purchase behavior enables a more customer-centric marketing strategy. Marketers can segment customers into various buckets based on the predicted LTV and, in turn, customize marketing messages or advertising copies to serve customers in different segments better. Furthermore, LTV predictions can directly inform marketing budget allocations and improve real-time targeting and bidding of ad impressions.

One challenge of LTV modeling is that some customers never come back, and the distribution of LTV can be heavy-tailed. The commonly used mean squared error (MSE) loss does not accommodate the significant fraction of zero value LTV from one-time purchasers and can be sensitive to extremely large LTV's from top spenders. In this article, we model the distribution of LTV given associated features as a mixture of zero point mass and lognormal distribution, which we refer to as the zero-inflated lognormal (ZILN) distribution. This modeling approach allows us to capture the churn probability and account for the heavy-tailedness nature of LTV at the same time. It also yields straightforward uncertainty quantification of the point prediction. The ZILN loss can be used in both linear models and deep neural networks (DNN). For model evaluation, we recommend the normalized Gini coefficient to quantify model discrimination and decile charts to assess model calibration. Empirically, we demonstrate the predictive performance of our proposed model on two real-world public datasets.

\end{abstract}

\section{Introduction}

There is a growing need for marketers to accurately predict a customer's future purchases in a long time horizon, such as one, two, or even five years. Such long term prediction is often called customer lifetime value (CLV or LTV). LTV predictions not only help the firm's financial planning but also inform marketing decisions and guide customer relationship management (CRM). With LTV predictions, it is straightforward to segment customers into various value buckets. Marketers can subsequently, decide how to improve the allocation of their marketing spend and determine the ideal target audiences for promotional offers, personalized customer messaging, exclusive deals, loyalty rewards programs, and ``white glove'' customer service treatment.

There is a body of literature on predicting the LTV of existing customers. Much of the developments evolve around the extension of the RFM (Recency, Frequency, Monetary Value) framework \citep{khajvand2011estimating}. The most well-known approach is the \emph{Buy Till You Die} (BTYD) family \citep{fader2005rfm, fader2009probability}. It is a probabilistic generative model for repeat purchases and customer churn. Both the customer churn and purchase behavior are assumed to follow some stochastic process. Multiple variants \citep{schmittlein1987counting, fader2005counting, fader2010customer} exist to either account for discrete-time purchase event data or reduce the computation burden.

In this paper, we focus on the LTV predictions of new customers, which has received far less attention. Predicting the LTV of new customers is essential to the advertising business. For example, marketers can treat the prediction as a Key Performance Indicator (KPI) and monitor it over time to continuously gauge the performance of customer acquisition marketing campaigns. The BTYD model family does not apply to new customers because it uses frequency and recency to differentiate customers. New customers, however, have identical purchase frequency and recency. The predictive signals must be extracted elsewhere - either the customer attributes obtained during customer sign-up or registration, or the product or service type of the initial purchase.

We approach the LTV prediction of new customers with supervised regression. Contrary to the BTYD model family, supervised regression leverages all customer-level features. It does not attempt to model the underlying dynamics of custom churn or repeat purchases but minimizes the specified prediction error instead. For the regression task, many standard machine learning methods are readily available, including linear regression, random forests, gradient boosting, support vector machines. We choose deep neural networks (DNN) as our workhorse due to its competitive performance and the ability to capture the complex and nonlinear relationships between predictive features and LTV.

It is relatively easy to predict aggregate business measures for financial planning. Accurately predicting the LTV of individual customers, however, is a far more difficult task. There are two main data challenges for this regression problem. The first is that many customers are one-time purchasers and never purchase again, resulting in many zero value labels. The second is that for returning customers, the LTV is volatile, and the distribution of LTV is highly skewed. A few high spenders could account for a significant fraction of the total customer spend, which embodies the spirit of the 80/20 rule.

Mean Squared Error (MSE), despite its dominant presence in regression modeling, is not the ideal choice for handling such data challenges in the context of LTV prediction. MSE ignores the fact that LTV labels are a mix of zero and continuous values and forces the model to learn the average of the two distributions. The squared term is also highly sensitive to outliers. Most large-scale training algorithms use stochastic gradient descent, noisy and occasionally exploding gradients computed from mini batches of training examples can easily cause numerical instability or convergence issues. We propose a mixture loss derived from the zero-inflated lognormal (ZILN) distribution. The loss handles the zero and extreme large LTV labels by design.

The DNN architecture, coupled with the ZILN loss, has several advantages compared with traditional regression models. First, it is capable of predicting the churn probability and LTV value simultaneously. It reduces the engineering complexity of building a two-stage model \citep{vanderveld2016engagement} --- a binary classification model to predict repeat purchase propensity, followed by a regression model to predict the LTV of returning customers predicted in stage 1. Second, it provides a full probabilistic distribution of LTV, and thus allows uncertainty quantification of point predictions.

For model evaluation, we propose using the normalized Gini coefficient to measure a model's ability to differentiate high-value customers from low-value ones. It is preferred over MSE due to its robustness to outliers and better business interpretation. We also suggest using decile charts to measure model calibration qualitatively.

The remainder of the paper is organized as follows. Section 2 briefly reviews related work. Section 3 presents the proposed DNN model along with the ZILN loss. We describe the normalized Gini coefficient and decile charts for model evaluation in Section 4 and demonstrate the proposed model empirically on several public-domain datasets. Finally, Section 5 concludes our discussion of LTV prediction models.

\section{Related Work}

\citet{gupta2006modeling} provide a comprehensive review of LTV methodologies. They present evidence that machine learning methods such as random forest \citep{breiman2001random} have superior performance than historically popular RFM and BTYD models because they can incorporate a variety of additional features.

\citet{vanderveld2016engagement, chamberlain2017customer} use a two-stage random forest model to predict the LTV of users of e-commerce sites. Stage one predicts purchase propensity --- a binary classification for whether or not the user is predicted to purchase for the specified time window. Stage two predicts the dollar value for users who were predicted to purchase in stage one. The two-stage approach is a natural way to build up the LTV prediction and provides insights into different factors that drive LTV. The main drawback is the added complexity to maintain two models.

An alternative two-stage approach is to build regression models for purchase frequency and average order value (or margin) separately, and then combine them into an LTV prediction model \citet{venkatesan2004customer}. This strategy can also be found in the RFM and BTYD framework. \citet{fader2005rfm} assume a Pareto/Negative Binomial Distribution (Pareto/NBD) for recency and frequency with purchase values following an independent Gamma/Gamma distribution. The decomposition, however, relies on a shaky assumption that purchase order value is independent of purchase frequency. In practice, for example, frequent purchasers may spend less on each purchase.

Many researchers prefer a direct approach for LTV prediction, which is more straightforward and often leads to higher prediction accuracy \citep{gupta2006modeling}. \citet{malthouse2005can} uses the LTV as the dependent variable in a regression model. The authors also consider various transformations of LTV, including Box-Cox transformation \citep{sakia1992box}, to stabilize the variance in the regression model, square-root, or logarithmic transformations to make the distribution of LTV much less right-skewed. The transformations, however, make the predictions biased by design. For example, due to Jensen's inequality, the exponential of the expectation of a logarithmically transformed variable is no greater than the expectation of the original variable.

\citet{benoit2009benefits} advocate a quantile regression approach that models the conditional quantiles of the response variable, such as the median, as opposed to the conditional mean modeling of standard least-squares regression. With standard mean regression techniques, a single point estimate of LTV is returned for each customer. The point estimate, however, does not contain information about the dispersion of observations around the predictive value. Prediction intervals can be obtained based on asymptotic normality, but quantile regression offers a more principled way of quantifying the uncertainty associated with the LTV prediction. For example, a 90\% prediction interval of LTV can be given by the 5th and 95th predicted percentile.

\citet{chamberlain2017customer} recognize the unusual distribution of the LTV. A large percentage of customers have an LTV of zero. Of the customers with positive LTV, the values differ by several orders of magnitude. The authors address this problem by modeling the percentile rank of the LTV and subsequently map them back to real values for use in downstream tasks. \citet{sifa2018customer} explain a similar problem in the context of LTV predictions for players of a free-to-play game. Only a small subset of users ever makes a purchase and drives the largest part of revenue. The authors suggest training a DNN with synthetic minority oversampling (SMOTE) \citep{chawla2002smote} to achieve better prediction performance. SMOTE is a data augmentation technique that creates synthetic entities of the minority class during the model training phase to regularize the prediction models and learn structures representing minority entities.

\citet{chamberlain2017customer} find that a DNN with enough hidden units can achieve comparable performance to Random Forest. The author also shows that for customer churn prediction, the wide-and-deep \citep{cheng2016wide} model yields further performance gains, because it combines the strengths of a wide linear model (for memorization) and a deep neural network (for generalization).

\section{DNN Model With ZILN Loss}

Regression labels are the total amount of customer spend in a fixed time horizon after the initial purchase. We exclude the first purchase value because our main interest is customers' future residual value. An exact number of years for the prediction horizon is preferred to avoid seasonal fluctuations. Practically, the prediction horizon is 1, 2, or 3 years. A longer-term model is often infeasible due to the length of historical data required to construct training labels. For example, both \citet{vanderveld2016engagement} and \citet{chamberlain2017customer} choose to predict a 1-year prediction horizon.

Regression features can be extracted from a variety of sources. Purchase history, when available, is often the primary source for feature engineering. Other common features include customer demographics, customer cohorts, return history, quality indicators of customer service. \citet{vanderveld2016engagement} use customer engagement levels before a final purchase decision to predict the LTV of an e-commerce site user. Such features include the number of opens and clicks for marketing emails, deal impressions, and searches. \citet{sifa2018customer} predict the LTV of players of a free-to-play game using activity-related metrics such as number of sessions, rounds and days played, amount of in-game currency purchased; temporal patterns of behavior, such as time between first and last session and inter-day and inter-session time distribution; meta-features such as country of origin, type of device, operating system, and customer acquisition channel. \citet{chamberlain2017customer} combine handcrafted features with unsupervised neural embedding learned from session and app logs of customer product views. The resulting model is both aware of domain knowledge and can learn rich patterns of customer behavior from raw data.

We consider DNN as our workhorse of LTV prediction for three reasons: performance, flexibility, and scalability. DNN has enjoyed recent successes in computer vision, speech recognition, recommendation systems, natural language processing, and many other areas. Evidence from its popularity in online data science competitions, DNN has a very competitive performance on tabular data due to its ability to capture the complex and nonlinear relationships between features and labels. DNN is also extremely flexible. One can easily customize its loss function, which makes it an ideal model for our ZILN loss. It can gracefully handle all types of features, including numerical, categorical, and even multivalent features. Sparse categorical features can be encoded as embedding and learned in a supervised way. Deep learning frameworks such as TensorFlow and Pytorch provide highly scalable implementations of DNN that is capable of handling very large datasets with millions or even billions of customers.

\begin{figure}
\centering
  \includegraphics[width=.7\linewidth]{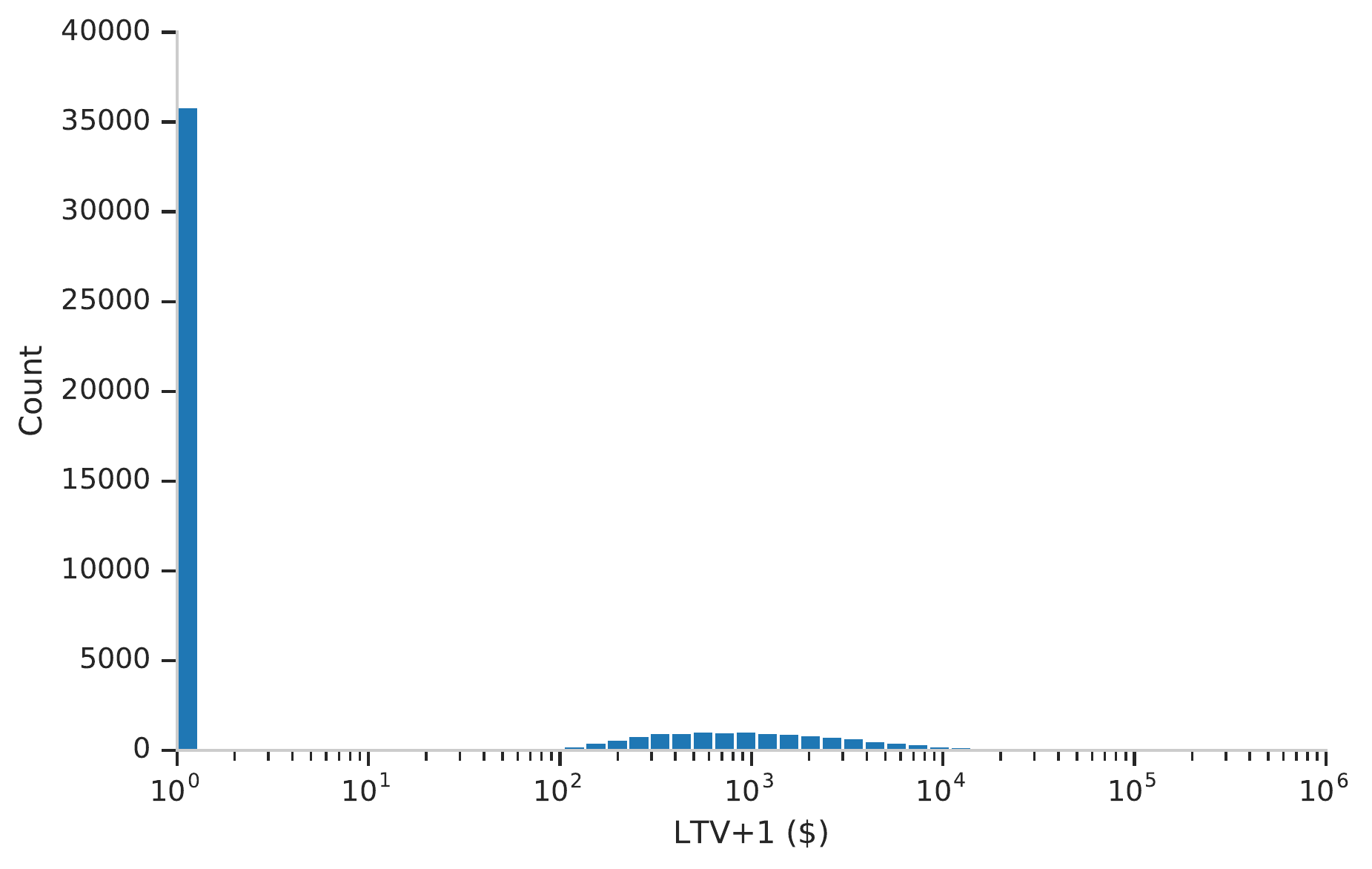}
  \caption{An illustration of a typical LTV distribution. A large proportion of
    customers are one-time purchasers. Returning customers' LTV can vary by
    orders of magitude.}
  \label{fig: ltv_hist}
\end{figure}

The distribution of LTV labels poses some challenges for the standard MSE regression loss. We show the LTV distribution of customers from a typical online advertiser in Figure~\ref{fig: ltv_hist}. The huge spike at value zero indicates the large fraction of one-time purchasers with zero LTV. For returning customers, the range of the LTV is also wide. The small set of high-value customers spent orders of magnitude more than a typical customer. The MSE can over-penalize prediction errors for high-value customers. Model training can also become unstable and sensitive to outliers. Swapping the MSE loss with quantile loss mitigates the outlier issue, but the model can no longer predict mean LTV, which is often desired.

We propose a mixture loss derived as the negative log-likelihood of a ZILN distribution. Such a mixture loss enables simultaneous learning of the purchase propensity and monetary value. The resulting model has half of the engineering complexity of a two-stage model --- typically a binary classification model to predict purchase propensity followed by a regression model to predict the monetary value for customers who are predicted to purchase \citep{vanderveld2016engagement}. The heavy-tailed lognormal distribution, which takes only positive values and has a long tail, is a natural choice for modeling the LTV distribution of returning customers. Mathematically, the lognormal loss, denoted as $L_\text{Lognormal}$, is derived as the negative log-likelihood of a lognormal random variable with mean $\mu$ and standard deviation parameter $\sigma$
\begin{equation}
  L_\text{Lognormal}(x; \mu, \sigma) = \log (x \sigma \sqrt{2 \pi}) +
  \frac{(\log x - \mu)^2}{2 \sigma^2}.
\end{equation}

It can be viewed as the weighted MSE on the log-transformed $X$, where the standard deviation parameter $\sigma$ plays the weighting role. Furthermore, the standard deviation parameter can also depend on the input features, just like the mean parameter, which implies a heteroscedastic lognormal distribution for LTV. Obtaining a good estimate of $\sigma$ is crucial as it directly influences the unbiasedness of the mean prediction due to the following formula
\begin{equation}\label{eq: mean of lognormal}
\E(X) = \exp\left(\mu + \frac{\sigma^2}{2}\right).
\end{equation}

\begin{figure}
  \centering
  \includegraphics[width=.7\linewidth]{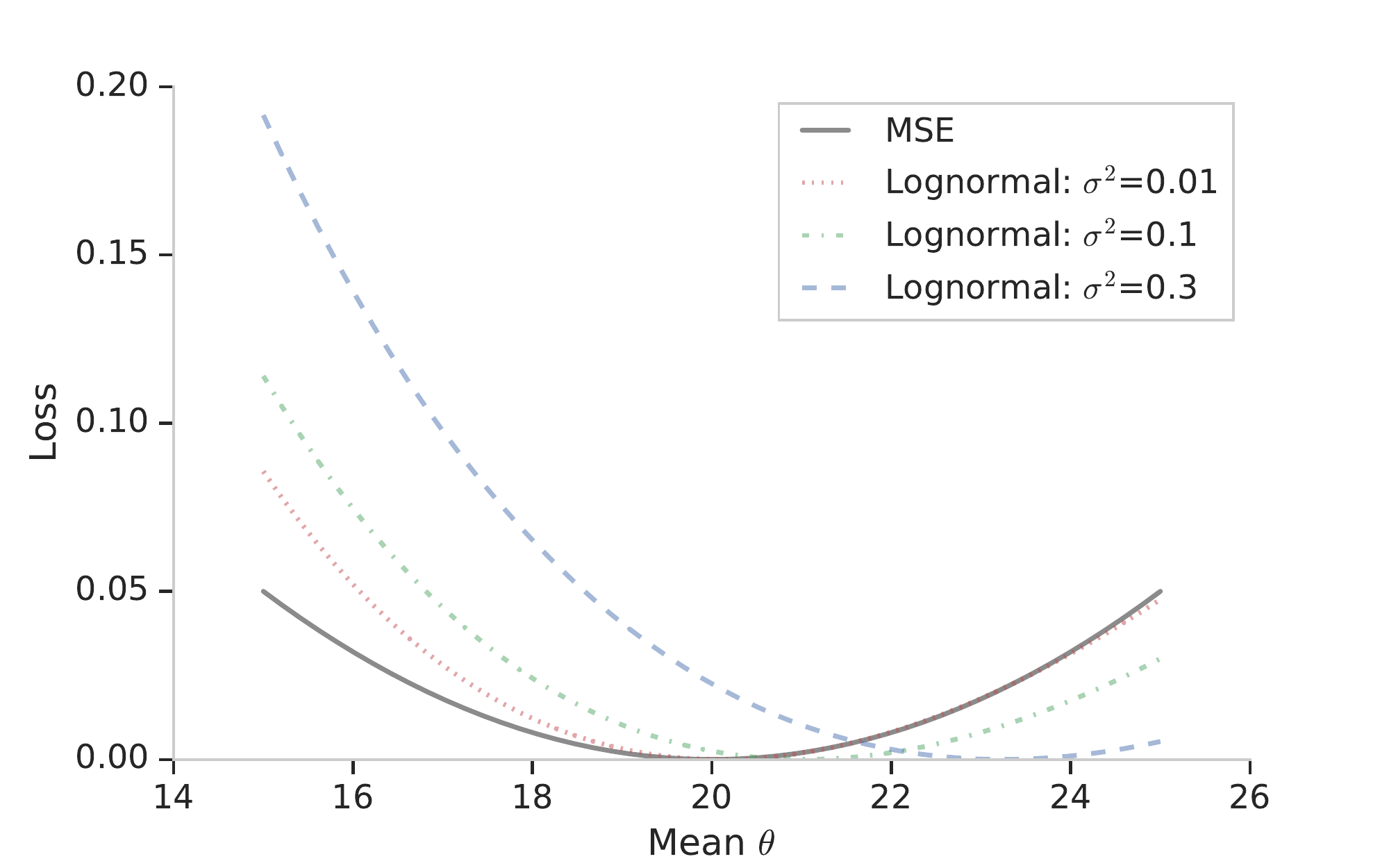}
  \caption{Compare the MSE loss to the lognormal loss as a function of the mean
    parameter $\theta$ with a single observation ($x=20$).}
  \label{fig:loss_function}
\end{figure}

We compare the MSE and the lognormal loss in Figure~\ref{fig:loss_function}. It shows that the MSE loss penalizes symmetrically around the observed value, while lognormal loss penalizes less on high values. The argmin increases as $\sigma$ increases.

The ZILN loss can be similarly derived as the negative log-likelihood of a ZILN distributed random variable with $p$ as the probability of being nonzero
\begin{equation}
L_\text{ZILN}(x; p, \mu, \sigma) = - \mathbbm{1}_{\{x = 0\}} \log (1 - p) -
\mathbbm{1}_{\{x > 0\}} (\log p - L_\text{Lognormal}(x; \mu, \sigma)),
\end{equation}
where $\mathbbm{1}$ denotes the indicator function.

The loss can be decomposed into two terms --- the first corresponding to the classification loss whether the customer is a returning customer, and the second corresponding to the regression loss of repeat customer's LTV.
\begin{equation}
L_\text{ZILN}(x; p, \mu, \sigma) = L_\text{CrossEntropy}(\mathbbm{1}_{\{x >
  0\}}; p) + \mathbbm{1}_{\{x > 0\}} L_\text{Lognormal}(x; \mu, \sigma).
\end{equation}

\begin{figure}
\centering
  \includegraphics[width=.6\linewidth]{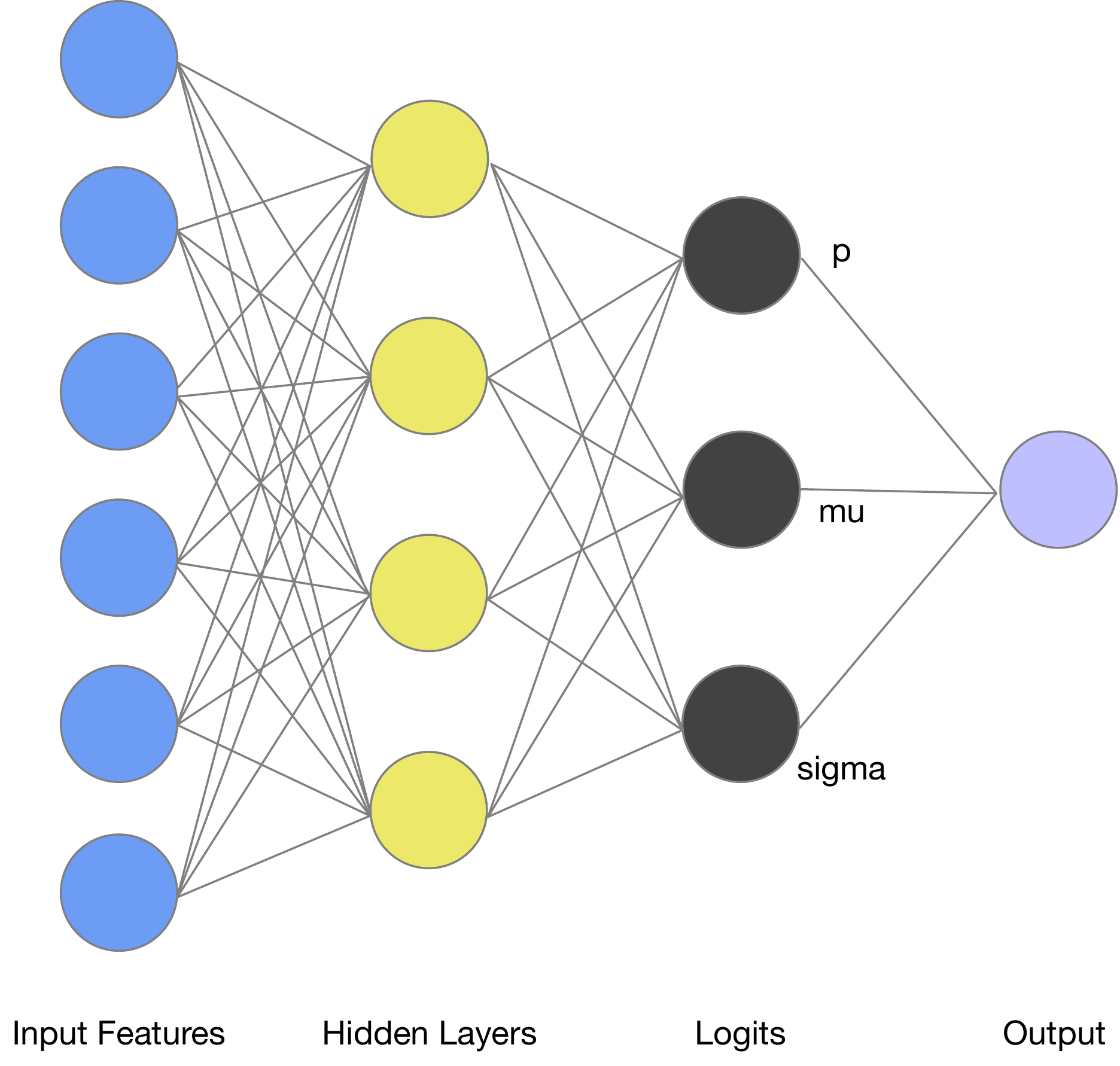}
  \caption{Network structure of DNN with the ZILN loss. $p$ represents the
    probability of returning customers; $\mu$ and $\sigma$ refer to the mean and
  standard deviation parameters of the lognormal distribution for the LTV of returning customers.}
  \label{fig: nn_structure}
\end{figure}

We present a visualization of the network in Figure~\ref{fig: nn_structure}. The last layer of the DNN has three pre-activation logits units, separately determining the returning purchase probability $p$, mean $\mu$, and standard deviation $\sigma$ of LTV for returning customers. The three activation functions are sigmoid, identity, and softplus, respectively. The middle layers of the DNN are essentially shared representations of two related tasks --- classification of returning customers and prediction of returning customer spend. This architecture encourages the model to generalize better on each task, which shares the core idea of multi-task learning \citep{ruder2017overview}.

Another key advantage of the ZILN loss is that it provides a full prediction distribution. We obtain not only the probability of returning but also the value distribution of LTV for returning customers. In addition to mean LTV prediction, the uncertainty of LTV predictions can be assessed using quantiles of a lognormal distribution as in general quantile regression.

\section{Evaluation Metrics}

For the binary classification problem of returning versus non-returning customers, standard classification metrics such as Area Under the Receiver Operating Curve (AUC) \citep{coussement2010improved, lemmens2006bagging} or Area Under the Precision-Recall Curve (AUC\_PR) \citep{boyd2013area} can be readily employed. AUC is a discriminative measure with a probabilistic interpretation. Given a randomly chosen returning customer and a randomly chosen non-returning customer, AUC is the probability that the classifier under evaluation can correctly predict the returning customer having a higher returning probability than the non-returning customer. AUC lies between 0.5 and 1. The closer the value is to 1, the better the classifier is at discriminating returning customers from non-returning customers.

For the regression problem of LTV prediction, commonly used measures such as (root) MSE or Mean Absolute Error (MAE) are less appropriate with our ZILN loss. Mean regression achieves the best predictive performance when the MSE is used as the training loss, while quantile regression excels when the MAE is considered the training objective. MSE, in particular, amplifies large prediction errors and tends to over-emphasizes high-value customers in training.

Traditionally, Pearson correlation \citep{donkers2007modeling, vanderveld2016engagement} between the actual and predicted LTV is used to assess prediction quality. The measure, however, can be sensitive to outliers in the data. \citet{chamberlain2017customer} use the Spearman rank correlation as a more robust alternative.

We evaluate the predictive performance of an LTV model from two aspects: discrimination and calibration. Model discrimination indicates the model's ability to differentiate high-value customers from the rest. Model calibration refers to the agreement between the actual and predicted LTV.

\subsection{Model Discrimination}

\citet{donkers2007modeling} propose the hit-rate measure, which is the percentage of customers whose predicted LTV falls into the same category as their true LTV. For example, if the 25\% most valuable customers have an LTV of more than 200, the hit rate then measures how many of these customers also have a predicted LTV of more than 200. \citet{malthouse2005can} also considers an ordering-based hit-rate. For the example above, it measures how many of the top 25\% customers based on actual LTV have a predicted LTV that is in the top 25\% of predicted LTV.

We consider a metric that generalizes the hit-rate but does not require the specification of the hit-rate level or percentile. Italian statistician and sociologist Corrado Gini proposed the Gini coefficient or Gini index \citep{gini1997concentration} over a century ago. It is frequently used in economics to measure the inequality of income or wealth distribution. The \emph{label} Gini coefficient can be computed in three steps.

\begin{enumerate}

\item Sort the true LTV in descending order (note that the original definition was to sort in ascending order, we change it for more straightforward interpretations of high-value customers).

\item Draw the Lorenz curve \citep{gastwirth1972estimation} which shows the cumulative percentage of total LTV ($y$-axis) against the cumulative percentage of customers ($x$-axis). A point $(x, y)$ on the curve means the top $x$-percentage of customers capture $y$-percentage of total customer value. When $(x, y) = (20, 80)$, it becomes the well-known 80/20 rule \citep{trueswell1969some}, aka the Pareto Principle.

\item The Gini coefficient is double of the area between the Lorenz Curve and the 45$\degree$ diagonal line, which corresponds to a random ordering of customers. It reflects the inequality of customer spending --- the larger the value, the more inequality of the distribution.

\end{enumerate}

\begin{figure}
\centering
  \includegraphics[width=0.7\linewidth]{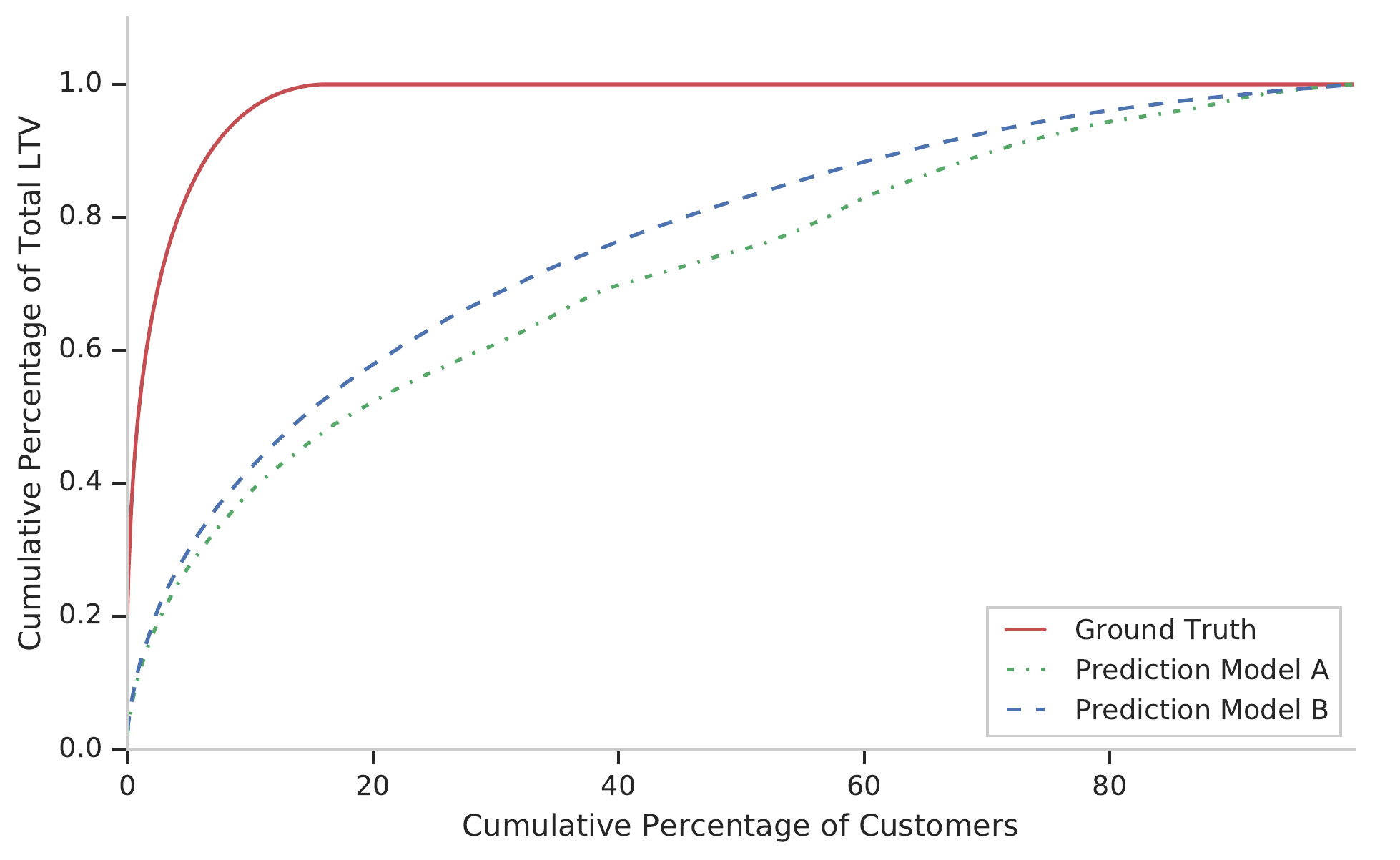}
  \caption{An illustration of gain chart. We compare two prediction models A and B. Each point $(x, y)$ on the gain curve denotes the $y$-percentage of total revenue is contributed by the predicted top $x$-percentage of customers. Model A is better at discriminating customers than model B. Ground truth refers to the Lorenz curve which is constructed by sorting customers by their true LTV.}
  \label{fig: gain_chart}
\end{figure}

We compute the \emph{model} Gini coefficient by substituting the true LTV with the predicted LTV in the sorting step 1. The resulting chart in step 2 is also known as the cumulative gain chart \citep{berry2004data}. We show a typical chart in Figure~\ref{fig: gain_chart} with the Lorenz curve (sorting by true LTV) and two model curves (sorting by predicted LTV). The closer the model curve is to the Lorenz curve, the better the model is at differentiating customers. The resulting model Gini coefficient resonates more with marketing professionals due to its interpretation and close resemblance to the 80/20 rule.

Similar to AUC, the model Gini coefficient is a discriminative measure. Gini equals two times AUC minus 1 for any binary classifier. The model Gini coefficient is purely based on the ranks of the predictions and is not sensitive to model miscalibration (to be discussed in the next section). It is especially useful when the use case is to segment customers based on predicted LTV.

The ratio between the model Gini coefficient and the label Gini coefficient yields the \emph{normalized} model Gini coefficient. It lies between 0 and 1, with the upper bound achieved by perfect LTV predictions and the lower bound corresponding to a random ordering of customers. Normalized Gini coefficient can be viewed as an extension to the hit-rate criterion but without the need to specifying the hit-rate level or percentile.

We compute a third type of Gini coefficient by replacing the true LTV with the first purchase value in step 1. We call it the \emph{baseline} Gini coefficient. The high correlation between the first purchase value and the LTV makes the baseline Gini coefficient a reasonable and practical lower bound of any model Gini coefficient. Further improvements of the baseline Gini coefficient can then be attributed to the addition of other predictive signals such as customer attributes, the metadata of the first purchase, and the non-purchase behavior before the first purchase.

\subsection{Model Calibration}

For binary classification problems, calibration plots \citep{cohen2004properties} have been widely adopted to evaluate soft classifiers that yield continuous probability predictions. A calibration plot is a goodness-of-fit diagnostic graph with the predicted probabilities on the x-axis, and the fraction of positive labels on the y-axis. For example, if we predict a 20\% probability of being a high-value customer, the observed frequency of high-value customers should be approximately 20 out of 100 customers with such a prediction. Perfect predictions should be on the 45-degree line.

For regression problems, the calibration plot becomes a simple scatter plot. When the labels have a highly skewed distribution, as in our LTV problem, the scatter plot may struggle to illustrate the calibration on small prediction regions. To improve the graphical presentation, we plot the labels by decile of predictions in a \emph{decile chart}, a close sibling of the cumulative gain chart and the lift chart \citep{berry2004data}. For each decile of predictions, we compare the average prediction and the average label side by side.

\begin{figure}
\centering
  \includegraphics[width=1\linewidth]{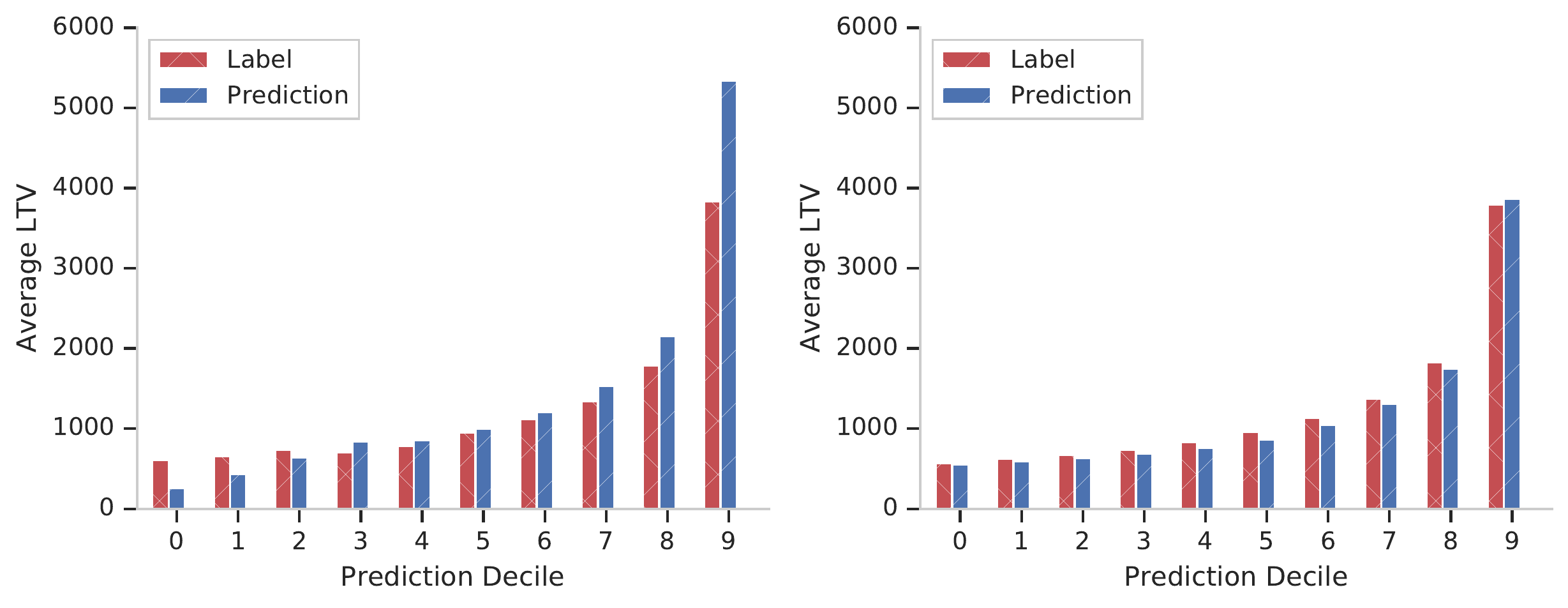}
  \caption{An illustration of the decile chart. The left panel shows a bad model calibration with over-prediction in high deciles and under-prediction in lower deciles. The right panel shows a good model calibration where the predicted LTV matches closely to the true LTV for each decile.}
  \label{fig:decile_chart}
\end{figure}

A well-calibrated model should have the prediction mean closely match to the label mean for each prediction decile. Figure~\ref{fig:decile_chart} shows examples of both bad and good model calibration.

Moreover, the decile chart provides a qualitative assessment of model discrimination. A better discriminating model has more spread between deciles than a poorly discriminating model.

Besides visual checks of the decile chart, we suggest the decile-level Mean Absolute Percentage Error (MAPE) as a quantitative measure of model calibration. Let $\hat{y}_i$ and $y_i$ denote the prediction and label mean for customers in the $i$-th prediction decile. MAPE is computed as
\begin{equation}
\text{MAPE} = \sum_{i=1}^{10} \frac{|\hat{y}_i - y_i|}{y_i}.
\end{equation}

\section{Data Experiments}

We use two public-domain datasets to evaluate the predictive performance of our proposed model.

\subsection{Kaggle Acquire Valued Shoppers Challenge}

The dataset for the \href{https://www.kaggle.com/c/acquire-valued-shoppers-challenge}{Kaggle Acquire Valued Shoppers Challenge} competition contains complete basket-level shopping history for 311K customers from 33K companies. We consider the task of predicting each customer's total purchase value in the next 12 months following the initial purchase. Model features include the initial purchase amount, the number of items purchased, as well as the store chain, product category, product brand, and product size measure of each individual purchased item.

We restrict our experiment to the top twenty companies based on customer count and focus on the cohort of customers who first purchased between 2012-03-01 and 2012-07-01. For each company, we randomly pick 80\% of customers for model training and use the remaining 20\% for model evaluation. We conduct our experiment along two axes: model architecture and loss. Both linear and DNN model are considered. The ZILN loss is compared to the MSE loss. We additionally report the binary classification results of returning customer prediction.

We implement our models using the TensorFlow framework. Following standard practice, for categorical features, we use one-hot encodings in linear models and embeddings in DNN. For DNN, we consider two hidden layers with 64 and 32 number of units, respectively. We train each model for up to 400 epochs with a batch size of 1,024 and the Adam optimizer \citep{kingma2014adam} with a learning rate of 2e-4. We also apply an early stopping rule to prevent overfitting.

\paragraph{Spearman's Correlation}
Spearman's correlation for each model is reported in Table~\ref{tab:kaggle_spearman}. The ZILN loss outperforms the MSE loss with Spearman's correlation being on average 23.9\% higher for the linear model and 48.0\% higher for DNN. For the ZILN loss, we see on average 2.2\% improvement of DNN over linear due to DNN's increased model flexibility and complexity.

\begin{table}[!h]
\begin{center}
\begin{tabular}{lcrrcrr}
\toprule
{\bf Model}    && \multicolumn{2}{c} {\bf DNN} && \multicolumn{2}{c} {\bf Linear}\\
\cmidrule{3-4} \cmidrule{6-7}
{\bf Loss}           && {\bf MSE}    & {\bf ZILN}   && {\bf MSE}    & {\bf ZILN} \\
\midrule
{\bf Company}  &&        &&        &        &        \\
10000     && 0.062 & \textbf{0.311} && 0.189 & 0.302 \\
101200010 && 0.351 & \textbf{0.423} && 0.294 & 0.413 \\
101410010 && 0.344 & \textbf{0.384} && 0.364 & 0.376 \\
101600010 && 0.325 & \textbf{0.424} && 0.376 & 0.418 \\
102100020 && 0.289 & \textbf{0.415} && 0.335 & 0.403 \\
102700020 && 0.210 & \textbf{0.313} && 0.269 & 0.308 \\
102840020 && 0.275 & \textbf{0.383} && 0.270 & 0.378 \\
103000030 && 0.294 & \textbf{0.340} && 0.294 & 0.335 \\
103338333 && 0.404 & \textbf{0.463} && 0.435 & 0.460 \\
103400030 && 0.206 & \textbf{0.295} && 0.261 & 0.287 \\
103600030 && 0.269 & \textbf{0.312} && 0.287 & 0.304 \\
103700030 && 0.301 & \textbf{0.397} && 0.233 & 0.390 \\
103800030 && 0.308 & \textbf{0.409} && 0.371 & 0.397 \\
104300040 && 0.274 & \textbf{0.380} && 0.327 & 0.371 \\
104400040 && 0.321 & \textbf{0.391} && 0.359 & 0.378 \\
104470040 && 0.275 & \textbf{0.321} && 0.248 & 0.312 \\
104900040 && 0.277 & \textbf{0.390} && 0.225 & 0.385 \\
105100050 && 0.259 & \textbf{0.335} && 0.257 & 0.331 \\
105150050 && 0.225 & \textbf{0.300} && 0.246 & 0.293 \\
107800070 && 0.263 & \textbf{0.330} && 0.285 & 0.324 \\
\bottomrule
\end{tabular}
\end{center}
\caption{Spearman's correlation between true and predicted LTV on the Kaggle Acquire Valued Shoppers Challenge dataset (higher is better).}
\label{tab:kaggle_spearman}
\end{table}

\paragraph{Model Discrimination}
Table~\ref{tab:kaggle_gini} summarizes the normalized Gini coefficient for the four models plus the baseline model, in which customers are ranked by the initial purchase value. Compared to the baseline model, DNN-MSE, DNN-ZILN, and linear-ZILN have an average relative improvement of 10.6\%, 23.1\%, and 21.3\%, respectively. On the other hand, the linear model with the MSE loss underperforms the baseline model in some cases, indicating convergence issues of the mini-batch training with the presence of outliers. The ZILN loss outperforms the MSE loss in both linear (28.6\% relative improvement) and DNN (11.4\% relative improvement). DNN with the ZILN loss achieves the best model discrimination due to its model flexibility and characterization of the label distribution.

\begin{table}[!h]
\begin{center}
\begin{tabular}{lcrcrrcrr}
\toprule
{\bf Model}    && {\bf Baseline} && \multicolumn{2}{c} {\bf DNN} && \multicolumn{2}{c} {\bf Linear}\\
\cmidrule{5-6} \cmidrule{8-9}
{\bf Loss}     &&                && {\bf MSE} & {\bf ZILN}       && {\bf MSE} & {\bf ZILN} \\
\midrule
{\bf Company}  &&                &&           &                  &&           & \\
10000     &  & 0.813 && 0.750 & \textbf{0.866} && 0.498 & 0.862 \\
101200010 &  & 0.639 && 0.628 & \textbf{0.700} && 0.451 & 0.687 \\
101410010 &  & 0.350 && 0.482 & \textbf{0.510} && 0.472 & 0.500 \\
101600010 &  & 0.376 && 0.404 & \textbf{0.480} && 0.372 & 0.472 \\
102100020 &  & 0.582 && 0.605 & \textbf{0.683} && 0.577 & 0.676 \\
102700020 &  & 0.256 && 0.348 & \textbf{0.409} && 0.366 & 0.394 \\
102840020 &  & 0.484 && 0.504 & \textbf{0.573} && 0.375 & 0.571 \\
103000030 &  & 0.347 && 0.414 & \textbf{0.437} && 0.397 & 0.424 \\
103338333 &  & 0.323 && 0.504 & \textbf{0.558} && 0.485 & 0.552 \\
103400030 &  & 0.544 && 0.574 & \textbf{0.610} && 0.587 & 0.595 \\
103600030 &  & 0.410 && 0.427 & \textbf{0.460} && 0.410 & 0.451 \\
103700030 &  & 0.467 && 0.451 & \textbf{0.540} && 0.293 & 0.533 \\
103800030 &  & 0.574 && 0.605 & \textbf{0.652} && 0.626 & 0.650 \\
104300040 &  & 0.448 && 0.460 & \textbf{0.533} && 0.452 & 0.529 \\
104400040 &  & 0.581 && 0.629 & \textbf{0.663} && 0.623 & 0.651 \\
104470040 &  & 0.535 && 0.600 & \textbf{0.623} && 0.517 & 0.617 \\
104900040 &  & 0.539 && 0.526 & \textbf{0.618} && 0.343 & 0.613 \\
105100050 &  & 0.330 && 0.389 & \textbf{0.447} && 0.253 & 0.441 \\
105150050 &  & 0.614 && 0.655 & \textbf{0.689} && 0.646 & 0.684 \\
107800070 &  & 0.442 && 0.441 & \textbf{0.498} && 0.417 & 0.497 \\
\bottomrule
\end{tabular}
\end{center}
\caption{Normalized Gini coefficient on the Kaggle Acquire Valued Shoppers Challenge dataset (higher is better).}
\label{tab:kaggle_gini}
\end{table}

\paragraph{Model Calibration}
The decile-level MAPE for all four models are reported in
Table~\ref{tab:kaggle_mape}. The ZILN loss leads to significantly reduced
decile-level MAPE than the MSE loss --- 60.0\% lower for linear and 68.9\% for
DNN. With the ZILN loss, DNN further reduces the decile-level MAPE over the
linear model (5.3\% lower).

\begin{table}[!h]
\begin{center}
\begin{tabular}{lcrrcrr}
\toprule
{\bf Model}    && \multicolumn{2}{c} {\bf DNN} && \multicolumn{2}{c} {\bf Linear}\\
\cmidrule{3-4} \cmidrule{6-7}
{\bf Loss}           && {\bf MSE}    & {\bf ZILN}   && {\bf MSE}    & {\bf ZILN} \\
\midrule
{\bf Company} &&        &        &&        &        \\
10000     &  & 0.653 & \textbf{0.182} && 0.456 & 0.187 \\
101200010 &  & 0.405 & \textbf{0.129} && 0.375 & 0.165 \\
101410010 &  & 0.249 & 0.057 && 0.265 & \textbf{0.055} \\
101600010 &  & 0.259 & \textbf{0.055} && 0.196 & 0.056 \\
102100020 &  & 0.272 & \textbf{0.122} && 0.223 & 0.132 \\
102700020 &  & 0.185 & \textbf{0.040} && 0.163 & 0.052 \\
102840020 &  & 0.350 & \textbf{0.118} && 0.260 & 0.130 \\
103000030 &  & 0.191 & \textbf{0.042} && 0.252 & 0.045 \\
103338333 &  & 0.208 & \textbf{0.073} && 0.348 & 0.075 \\
103400030 &  & 0.265 & \textbf{0.143} && 0.212 & 0.148 \\
103600030 &  & 0.408 & \textbf{0.057} && 0.272 & 0.067 \\
103700030 &  & 0.280 & \textbf{0.115} && 0.289 & \textbf{0.115} \\
103800030 &  & 0.356 & 0.124 && 0.352 & \textbf{0.119} \\
104300040 &  & 0.392 & 0.087 && 0.138 & \textbf{0.082} \\
104400040 &  & 0.242 & 0.100 && 0.365 & \textbf{0.098} \\
104470040 &  & 0.253 & \textbf{0.084} && 0.231 & 0.099 \\
104900040 &  & 0.337 & 0.126 && 0.378 & \textbf{0.122} \\
105100050 &  & 0.328 & 0.096 && 0.136 & \textbf{0.091} \\
105150050 &  & 0.314 & \textbf{0.095} && 0.181 & 0.097 \\
107800070 &  & 0.371 & \textbf{0.087} && 0.305 & 0.109 \\
\bottomrule
\end{tabular}
\end{center}
\caption{Decile-level Mean Absolute Percentage Error (MAPE) on the Kaggle Acquire Valued Shoppers Challenge dataset (lower is better).}
\label{tab:kaggle_mape}
\end{table}

\paragraph{Returning Customer Prediction}
We also report the Area Under Precision-Recall Curve (AUC\_PR) for the binary classification task of returning customer prediction in Table~\ref{tab:kaggle_auc-pr}. The ZILN has a comparable performance the standard Binary Cross Entropy (BCE) loss.

\begin{table}[!h]
\begin{center}
\begin{tabular}{lcrrcrr}
\toprule
{\bf Model}    && \multicolumn{2}{c} {\bf DNN} && \multicolumn{2}{c} {\bf Linear}\\
\cmidrule{3-4} \cmidrule{6-7}
{\bf Loss}           && {\bf BCE}    & {\bf ZILN}   && {\bf BCE}    & {\bf ZILN} \\
\midrule
{\bf Company} &&        &        &&        &        \\
10000     &  & \textbf{0.911} & 0.910 &  & 0.906 & 0.907 \\
101200010 &  & \textbf{0.889} & \textbf{0.889} &  & 0.886 & 0.886 \\
101410010 &  & \textbf{0.878} & 0.877 &  & 0.876 & 0.876 \\
101600010 &  & \textbf{0.958} & \textbf{0.958} &  & 0.957 & 0.957 \\
102100020 &  & \textbf{0.968} & \textbf{0.968} &  & 0.967 & 0.966 \\
102700020 &  & \textbf{0.869} & 0.868 &  & 0.867 & 0.868 \\
102840020 &  & \textbf{0.958} & 0.957 &  & 0.957 & 0.957 \\
103000030 &  & 0.875 & \textbf{0.876} &  & 0.875 & 0.875 \\
103338333 &  & \textbf{0.965} & \textbf{0.965} &  & \textbf{0.965} & \textbf{0.965} \\
103400030 &  & \textbf{0.902} & \textbf{0.902} &  & \textbf{0.902} & 0.901 \\
103600030 &  & 0.836 & \textbf{0.837} &  & 0.834 & 0.834 \\
103700030 &  & 0.962 & \textbf{0.963} &  & 0.962 & 0.962 \\
103800030 &  & 0.936 & \textbf{0.937} &  & 0.935 & 0.935 \\
104300040 &  & \textbf{0.925} & \textbf{0.925} &  & 0.923 & 0.923 \\
104400040 &  & 0.942 & \textbf{0.943} &  & 0.941 & 0.941 \\
104470040 &  & \textbf{0.889} & 0.888 &  & 0.887 & 0.887 \\
104900040 &  & 0.907 & \textbf{0.908} &  & 0.906 & 0.906 \\
105100050 &  & \textbf{0.941} & \textbf{0.941} &  & 0.940 & 0.940 \\
105150050 &  & 0.871 & \textbf{0.872} &  & 0.871 & 0.871 \\
107800070 &  & \textbf{0.858} & \textbf{0.858} &  & 0.856 & 0.856 \\
\bottomrule
\end{tabular}
\end{center}
\caption{Area Under Precision-Recall Curve (AUC\_PR) for returning customer prediction on the Kaggle Acquire Valued Shoppers Challenge dataset (higher is better).}
\label{tab:kaggle_auc-pr}
\end{table}

\subsection{KDD Cup 1998}

The \href{https://kdd.ics.uci.edu/databases/kddcup98/kddcup98.html}{Second International Knowledge Discovery and Data Mining Tools Competition} (a.k.a., the KDD Cup 1998) provides a dataset collected by Paralyzed Veterans of America (PVA), a non-profit organization that provides programs and services for US veterans with spinal cord injuries or disease. The organization raised money via direct mailing campaigns and was interested in lapsed donors: people who have stopped donating for at least 12 months. The provided dataset contains around 200K such donors who received the 1997 mailing and did not make a donation in the previous 12 months. We tackle the same task of the competition, which is to predict the donation dollar value to the 1997 mailing campaign.

\begin{figure}
\centering
  \includegraphics[width=.7\linewidth]{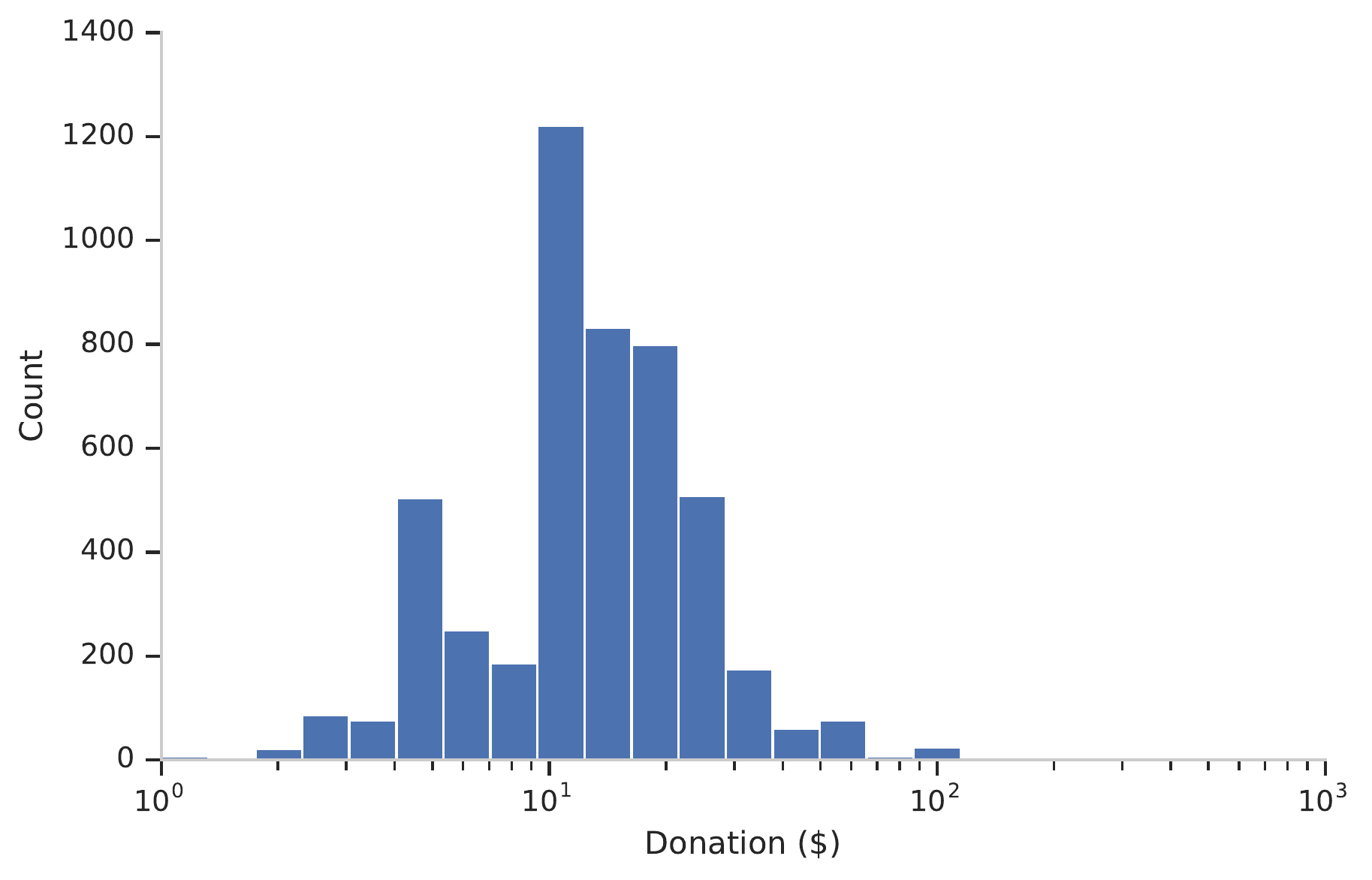}
  \caption{The distribution of the donation values on a log scale from the 5\% lapsed donors who responded to the 1997 mailing campaign.}
  \label{fig: donation_hist}
\end{figure}

The labels include a mix of zero and positive donation values. Around 95\% lapsed donors did not respond to the 1997 mailing campaign, thus assigned a zero label value. For the remaining 5\% lapsed donors, the distribution of the positive donation values is shown on a log scale in Figure~\ref{fig: donation_hist}. For the simplicity of the experiment, we fix DNN (four layers) as the model architecture and compare the ZILN loss to the MSE loss. We use a subset of available features such as donor demographics, promotion, and donation history. Due to the variations of the trained model over multiple runs with the same hyperparameters, we train each model 50 times and report the mean of the evaluation measures.

Compared to the MSE loss, ZILN loss leads to a higher Spearman's rank correlation (0.027 vs. 0.020). For model discrimination, the ZILN loss achieves a higher normalized Gini coefficient (0.190 vs. 0.184). The ZILN loss also outperforms the MSE for model calibration, with a smaller decile-level MAPE (0.176 vs. 0.210). The original objective was to maximize the total profit of the 1997 mailing campaign.

Each promotion mail has a cost of \$0.68. The total profit is calculated as the total of donation subtracted by the cost for donors with an expected revenue higher than \$0.68. The winner of the competition reported a total profit of \$14,712.24. Our best performing DNN model with the ZILN loss (among the 50 runs) achieves a total profit of \$15,498.24, representing a further 5\% relative increase.

\begin{figure}
\centering
  \includegraphics[width=1\linewidth]{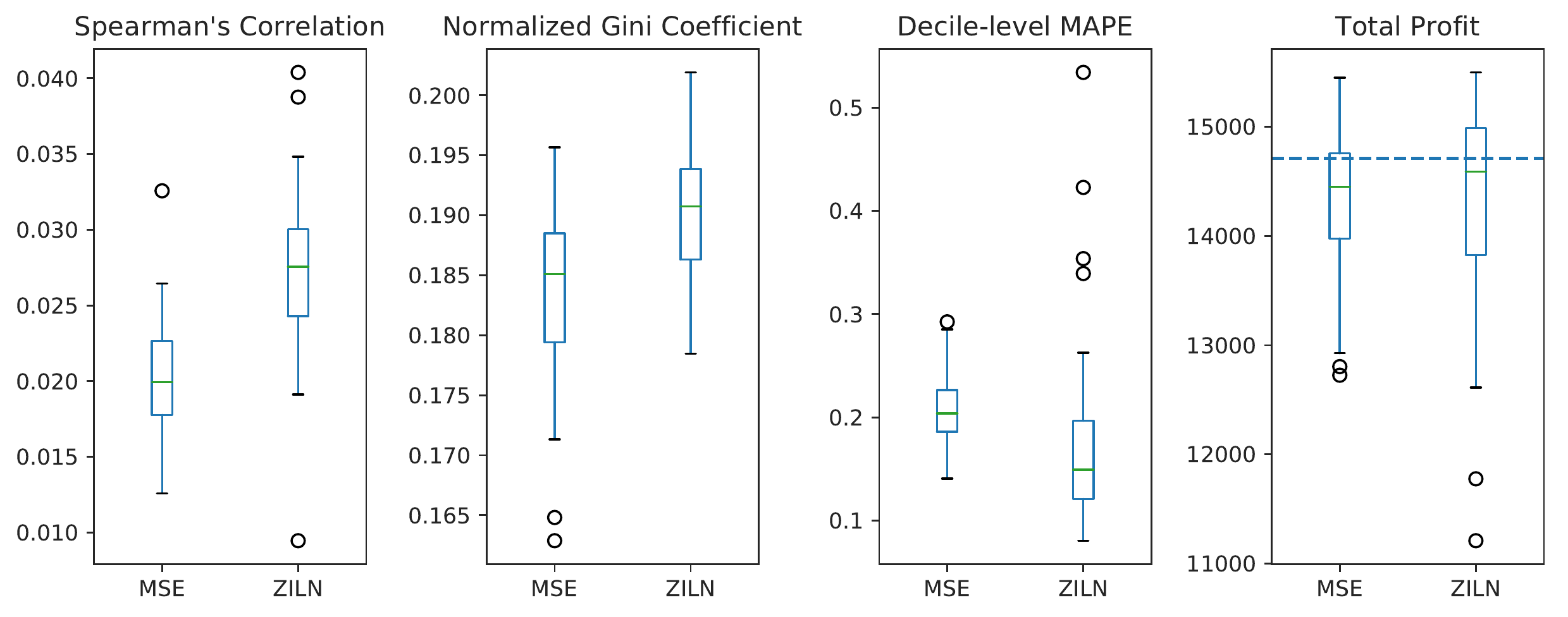}
  \caption{Predictive performance on the KDD Cup 1998 dataset. The boxplots compare the distribution of Spearman's correlation, normalized Gini coefficient, decile-level MAPE, and total profit between the MSE and ZILN loss over 50 repeat runs. The horizon line in the rightmost panel represents the total profit reported by the winner of the competition.}
  \label{fig:kdd_cup_98_eval}
\end{figure}

\section{Conclusion}

We have reviewed how LTV predictions can inform various marketing decisions. We use DNN to predict LTV of new customers based on customer attributes and purchase metadata. Our proposed mixture ZILN loss is tailored for the LTV label data, which is a mixture of zero and heavy-tailed values. We advocate the use of normalized Gini coefficients to quantify model discrimination and promote decile charts to assess model calibration. We demonstrate the competitive performance of our proposed method on two public datasets.

\subsubsection*{Acknowledgments}
The authors thank Jim Koehler, Tim Au, Georg Goerg, Dustin Tseng, Yael Grossman Levy, Henry Tappen for useful discussions, and Google Ads engineering and product team for their support. Special thanks go to our reviewers Jim Koehler, Nicolas Remy, David Chan, and Charis Kountouris for providing improvements to the original manuscript.

\clearpage

\bibliography{references}
\bibliographystyle{iclr2019_conference}

\clearpage

\begin{appendices}

\section{Relative efficiency in mean estimation of lognormal distribution}\label{app: efficiency}

We show that when the LTV of returning customers is truly lognormal distributed, the lognormal loss is more efficient than the MSE loss in estimating the mean LTV. Assume that $X_1, ..., X_n$ are i.i.d. lognormal distributed with underlying normal distribution of mean and standard deviation parameter $\mu$ and $\sigma$, i.e., $Y_i = \log{X_i}$ normally distributed as $N(\mu, \sigma^2)$.

According to Equation \ref{eq: mean of lognormal}, the maximum likelihood estimator (MLE) of $\theta$ is
\begin{equation}
  \hat{\theta}_{MLE} = \exp\left(\bar{Y} + \frac{1}{2n}S_Y^2\right),
  \label{eq: mle}
\end{equation}
where $\bar{Y} = \sum_{i=1}^n Y_i / n$ and $S_Y^2 = \sum_{i=1}^n (Y_i -
\bar{Y})^2$.

When the sample size $n$ is large, \citet{finney1941distribution} showed the variance of $\hat{\theta}$ can be approximated as
\begin{equation}
\Var\left(\hat{\theta}_{MLE}\right) \approx \frac{1}{n}\left(\sigma^2 +
\frac{1}{2} \sigma^4\right)\exp\left(2\mu + \sigma^2\right).
\end{equation}

Alternatively, the MSE loss implies an arithmetic mean estimator of $\theta$:
\begin{equation}
  \hat{\theta}_{AVG}=\frac{1}{n}\sum_{i=1}^nX_i
  \label{eq: sample mean estimator}
\end{equation}

With the variance formula derived by \citet{johnson1994lognormal}
\begin{equation}
\Var(X_i) = \left(\exp\left(\sigma^2\right) -1\right) \exp(2\mu+\sigma^2),
\end{equation}
and the independence assumption of $X_i$'s, we have
\begin{equation}
\Var(\hat{\theta}_{AVG}) = \frac{1}{n}\left(\exp\left(\sigma^2\right)
-1\right)\exp\left(2\mu+\sigma^2\right).
\end{equation}

$\hat{\theta}_{AVG}$ has a larger variance than $\hat{\theta}_{MLE}$ since the latter includes only the first two terms of the Taylor expansion of $\exp(\sigma^2) - 1$. The relative efficiency between the two estimators becomes even larger when the underlying standard deviation parameter $\sigma$ is large.

The above efficiency comparison results can be extended and generalized to ZILN loss because in loss computation we use lognormal loss for positive labels and Binary Cross Entropy (BCE) loss for zero labels.

We replicate the simulations 2,000 times for each $\sigma$ value, where a random sample of size 10,000 are drawn from $\lognormal(0,\sigma^2)$ and randomly splitted into into equally sized training and testing sets. Three estimators of the mean were considered, including simple average (Equation \ref{eq: sample mean estimator}), maximum likelihood estimator (Equation \ref{eq: mle}), and an approximation of an unbiased estimator \citep{finney1941distribution}:
\begin{equation}
\hat{\theta}_{Finney} = \exp{\left(\hat{\mu} + \frac{1}{2}\hat{\sigma}^2\right)}
\left\{ 1 - \frac{\hat{\sigma}^2(\hat{\sigma}^2+2)}{4n^2} +
\frac{\hat{\sigma}^4(3\hat{\sigma}^4 + 44\hat{\sigma}^2 + 84)}{96n^2}\right\}.
\end{equation}

\begin{figure}
\centering
  \includegraphics[width=.7\linewidth]{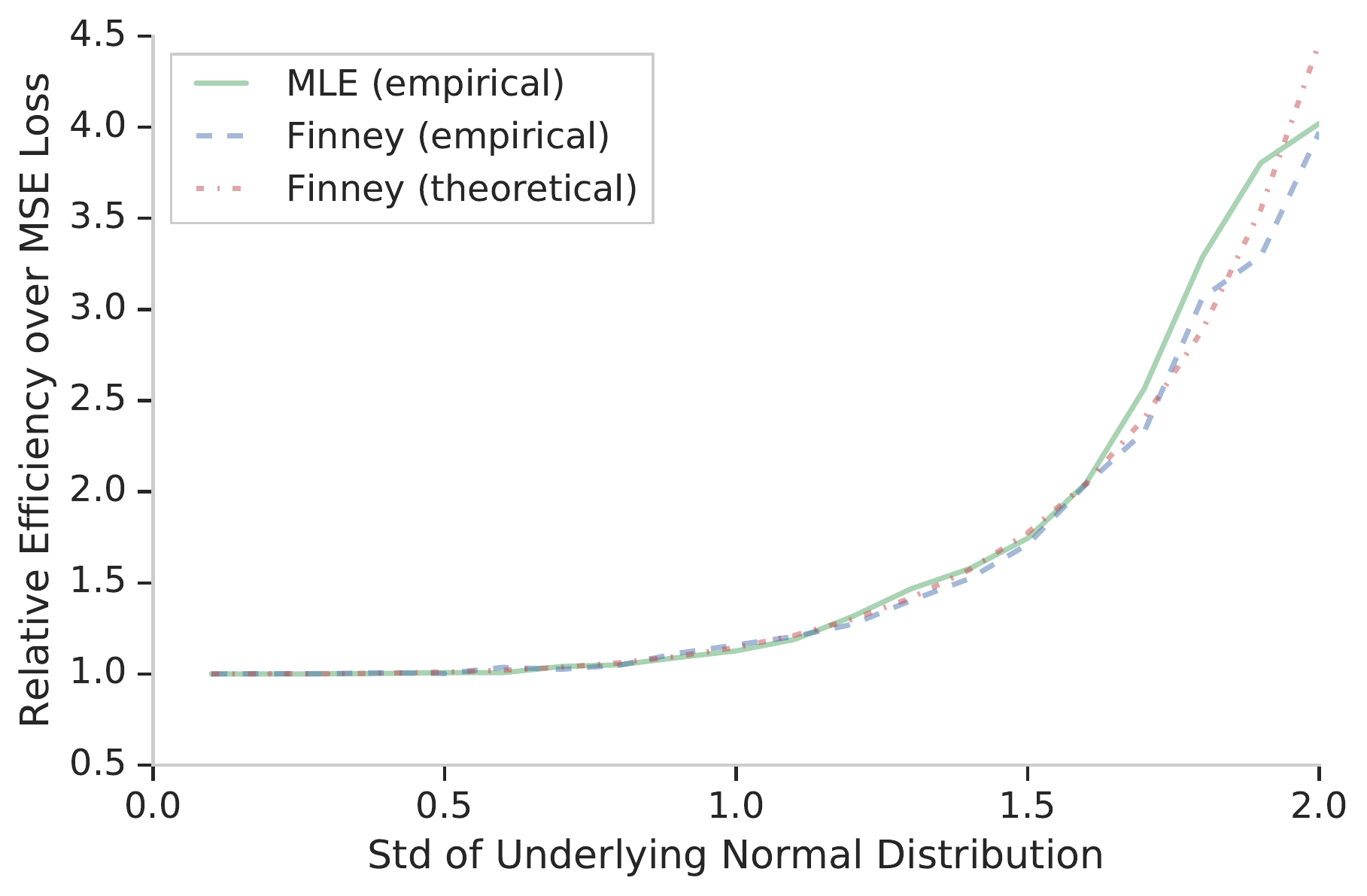}
  \caption{Relative efficiency of $\hat{\theta}_{MLE}$ (empirical) and $\hat{\theta}_{Finney}$ (empirical and theoretical) to the MSE estimator $\hat{\theta}_{AVG}$.}
  \label{fig: relative_efficiency_with_std}
\end{figure}

The three estimators are compared in terms of the mean squared error (MSE) on the testing sets, where the relative efficiency of $\hat{\theta}_{MLE}$ (Equation \ref{eq: mle}) and $\hat{\theta}_{Finney}$ are calculated as $MSE_{AVG}/MSE_{MLE}$ and $MSE_{AVG}/MSE_{Finney}$, respectively. The larger the relative efficiency, the better the estimator. The empirical estimates of the relative efficiency are shown in Figure \ref{fig: relative_efficiency_with_std}. $\hat{\theta}_{MLE}$ achieves a slightly higher relative efficiency than $\hat{\theta}_{Finney}$ for larger $\sigma$ values.

Figure \ref{fig: relative_efficiency_with_std} also shows the theoretical relative efficiency of $\hat{\theta}_{Finney}$ derived by \citet{finney1941distribution} (Equation 27). The theoretical curve aligns well with the empirical curve, which validates our simulation study.
\begin{equation}
\frac{MSE_{AVG}}{MSE_{Finney}} = \left\{{\sigma}^2+
\frac{{\sigma}^4}{2}+\frac{1}{2n}\left({\sigma}^6+\frac{{\sigma}^8}
     {4}\right)\right\}/\left(\exp(\sigma^2)-1\right).
\end{equation}

\end{appendices}

\end{document}